\documentclass[pdflatex,sn-nature]{sn-jnl}

\usepackage[]{graphicx}%
\usepackage{multirow}%
\usepackage{amsmath,amssymb,amsfonts}%
\usepackage{amsthm}%
\usepackage{mathrsfs}%
\usepackage[title]{appendix}%
\usepackage{xcolor}%
\usepackage{textcomp}%
\usepackage{manyfoot}%
\usepackage{booktabs}%
\usepackage{algorithm}%
\usepackage{algorithmicx}%
\usepackage{algpseudocode}%
\usepackage{listings}%
\usepackage{comment} 

\theoremstyle{thmstyleone}%
\theoremstyle{thmstyletwo}%

\theoremstyle{thmstylethree}%

\begin{document}

\title[Article Title]{Experimental Demonstration of Beam-Driven Wakefield Acceleration in Laser-Plasma Filament}

\author*[1]{\fnm{M.} \sur{Galletti}}\email{mario.galletti@lnf.infn.it}

\author[1]{\fnm{L.} \sur{Verra}}

\author[1]{\fnm{A.} \sur{Biagioni}}

\author[1]{\fnm{M.} \sur{Carillo}}

\author[1]{\fnm{L.} \sur{Crincoli}}

\author[1]{\fnm{R.} \sur{Demitra}}

\author[1]{\fnm{G.} \sur{Parise}}

\author[1]{\fnm{G.} \sur{Di Pirro}}

\author[1]{\fnm{R.} \sur{Pompili}}

\author[1,2]{\fnm{F.} \sur{Stocchi}}

\author[1]{\fnm{F.} \sur{Villa}}

\author[3]{\fnm{A.} \sur{Zigler}}

\author[1]{\fnm{M.} \sur{Ferrario}}

\affil*[1]{INFN-LNF, Via Enrico Fermi 54, 00044 Frascati, Rome, Italy}
\affil[2]{Università di Roma Tor Vergata, Department of Physics,Via della Ricerca Scientifica 1, 00133 Rome, Italy}
\affil[3]{Racah Institute of Physics, Hebrew University, 91904 Jerusalem, Israel}



\abstract{Self-guided femtosecond laser pulses propagating in low-pressure gas can generate plasma filaments, establishing a new framework for plasma wakefield acceleration.
Unlike conventional schemes relying on mechanically confined or preformed plasma channels, this method exploits the intrinsic non-linear light-matter interaction, greatly reducing the energy required to generate plasma. This, in turn, allows to realise tunable stages, potentially operating above kHz repetition rate and with meter-scale interaction lengths and transverse sizes down to a few tens of micrometres.
Moreover, the laser-plasma filament reproducibility is intrinsically higher than state-of-the-art discharge-plasmas, where the breakdown process is initiated in a stochastic and uncontrolled manner. As a result, laser-based plasma formation offers improved reliability and control over plasma parameters.
Here we report a proof-of-principle experimental demonstration of beam-driven wakefield acceleration of electron bunches with an accelerating field exceeding 250\,MV/m in a laser-generated plasma filament.
The results are cross-checked with numerical simulation, showing an excellent agreement and providing a complete picture of the physical process.
Beyond particle acceleration, the concept bridges laser filamentation physics, advanced plasma photonics and compact accelerator technologies, offering a promising route towards sustainable, high-repetition-rate plasma-based facilities.}

\keywords{Laser filamentation, Non-linear light–matter interaction, Plasma-based electron acceleration, Compact accelerator technologies}



\maketitle

\par The propagation of ultrashort, high-intensity laser pulses in gases is governed by a dynamic balance between self-focusing and defocusing due to low-density plasma generated via multi-photon or tunnel ionisation~\cite{pap7:13,pap7:16}. 
This balance leads to intensity clamping~\cite{pap7:17,pap7:19} and to the formation of extended plasma channels~\cite{pap7:1,pap7:21}, commonly referred to as \textit{filaments}. 
Under suitable conditions, filaments can extend over tens of meters~\cite{pap7:22, papeer2019towards} and have been observed over distances approaching two kilometres in atmospheric propagation~\cite{pap7:24}.

\par Amongst numerous applications~\cite{pap7:25,kasparian2003white,pap6:8}, the adoption of laser-plasma filaments as stages for plasma wakefield acceleration (PWFA)~\cite{TAJIMA:1979} represents a new potential direction. 
When a charged particle bunch (the $driver$) travels in plasma, the plasma electrons move to shield the electromagnetic fields of the bunch~\cite{VERRA:2024}, resulting in a plasma electron density oscillation behind the bunch. 
The plasma density wake sustains longitudinal and transverse \textit{wakefields}~\cite{CHEN:1985}, that can be used to accelerate a trailing bunch (the $witness$) to high energies over short distances~\cite{KALLOS:2008b, litos2014high}, with quality suitable for applications such as radiation generation~\cite{POMPILI:2022,GALLETTI:2022a,GALLETTI:2024}.

\par Conventionally, plasmas for PWFA are generated by ionising gas through either high-current discharge~\cite{LEEMANS:2006}, field-ionisation induced by relativistic charged particle bunches~\cite{OCONNEL:2006}, or multiphoton-tunnel ionisation~\cite{DEMETER:2021}. 
However, the large amount of power required by these methods limits the repetition rates.
In the case of discharge or field ionisation, the parameters of plasma depend directly on the high-voltage discharge system or beam parameters, which may jitter substantially from event to event. 
The methods mentioned above relying on laser-induced ionisation, and further ones~\cite{pap9_Milch,pap32} adopted in laser-driven wakefield acceleration~\cite{Esarey2009laserdriven}, do not involve self-focusing, thus filamentation does not play a role in the generation of plasma.

\par In the case of plasma filaments, the laser self-focusing allows for depositing much less power into the system, decreasing the heat load. This, in turn, allows to increase in the achievable repetition rate. Moreover, the geometric properties of the filament can be adjusted by tuning the parameters of the laser pulse and the gas pressure~\cite{GALLETTI:2025}. 

\par In this paper, we report a proof-of-principle experimental demonstration of the use of laser-plasma filaments for PWFA, leading to consistent ($\sim 95\%$ successful acceleration events) and high-gradient ($>250\,$MeV/m) acceleration.
The plasma evolution is characterised by analysing the recombination light, showing an excellent agreement with numerical simulations. 

\paragraph{Filamentation Physics} \label{sec1}

\par Filamentation of high-power femtosecond laser pulses involves several non-linear effects such as diffraction, self-focusing, group-velocity dispersion, as well as plasma generation and energy losses due to multi-photon and tunnel ionisation~\cite{pap3,pap6}. 
To investigate the phenomenon, we adopt a theoretical model based on the non-linear wave equation using the slowly varying envelope approximation~\cite{pap3,pap5,pap7:15,pap7:37,pap7:38} describing the electric field A$\equiv$ A(x, y, z, t) as~\cite{GALLETTI:2025}:
\newpage
\begin{multline}
\frac{\partial{A}}{\partial{z}}=\frac{i}{2k}\left(\frac{\partial^2{}}{\partial{x^2}}+\frac{\partial^2{}}{\partial{y^2}}\right)A-\frac{ik''}{2}\frac{\partial^2{}}{\partial{t^2}}A \\ 
+i k_0 n_2 \left( (1-f)I + f  \int_{-\infty}^{t} R(t-t') I(t') dt' \right) A \\ 
-\frac{\sigma}{2}(1+i\omega_0\tau_c)n_{pe} A-\frac{\beta_K}{2} I^{K-1} A,
\label{envelope}
\end{multline}
where $I=|A(x, y, z, t)|^2$ is the intensity of the laser pulse, $\omega_0$ and $k$  the laser frequency and wave number.
On the right-hand side, the first term accounts for diffraction within the transverse planes; the second term for the group velocity dispersion (GVD) of an ultra-short laser pulse, where $k''$ (in units of fs$^2$/cm) is the GVD coefficient; 
the third for the Kerr effect~\cite{pap8, sprangle2004ultrashort,pap3,pap5,pap7:15,pap7:37} with the instantaneous and delayed components in parentheses, where $k_0=\omega_0\tau_c/n_0$ ($\tau_c$ the characteristic time for electron-neutral inverse Bremsstrahlung), $n_0$ is the linear index of refraction, \textit{f}$= 0.5$ the coefficient due to stimulated molecular Raman scattering~\cite{pap3,pap7}, R(t) the exponential function with a characteristic time of about 70\,fs, and n$_2$ the non-linear index of refraction; the fourth term for plasma absorption and defocusing where $\sigma$ is the cross-section for the inverse Bremsstrahlung process following the Drude model~\cite{pap5:29}, and $n_{pe}\equiv n_{pe}$(x, y, z, t) the electron density of the plasma generated by ionisation; finally, the last term accounts for energy absorption due to ionisation where $\beta_K$ is the multi-photon ionisation rate parameter~\cite{pap7:37,pap7:41}, and K the number of photons needed to ionize each atom or molecule once. 

\begin{figure}[ht]
\centering
\includegraphics[width=\textwidth]{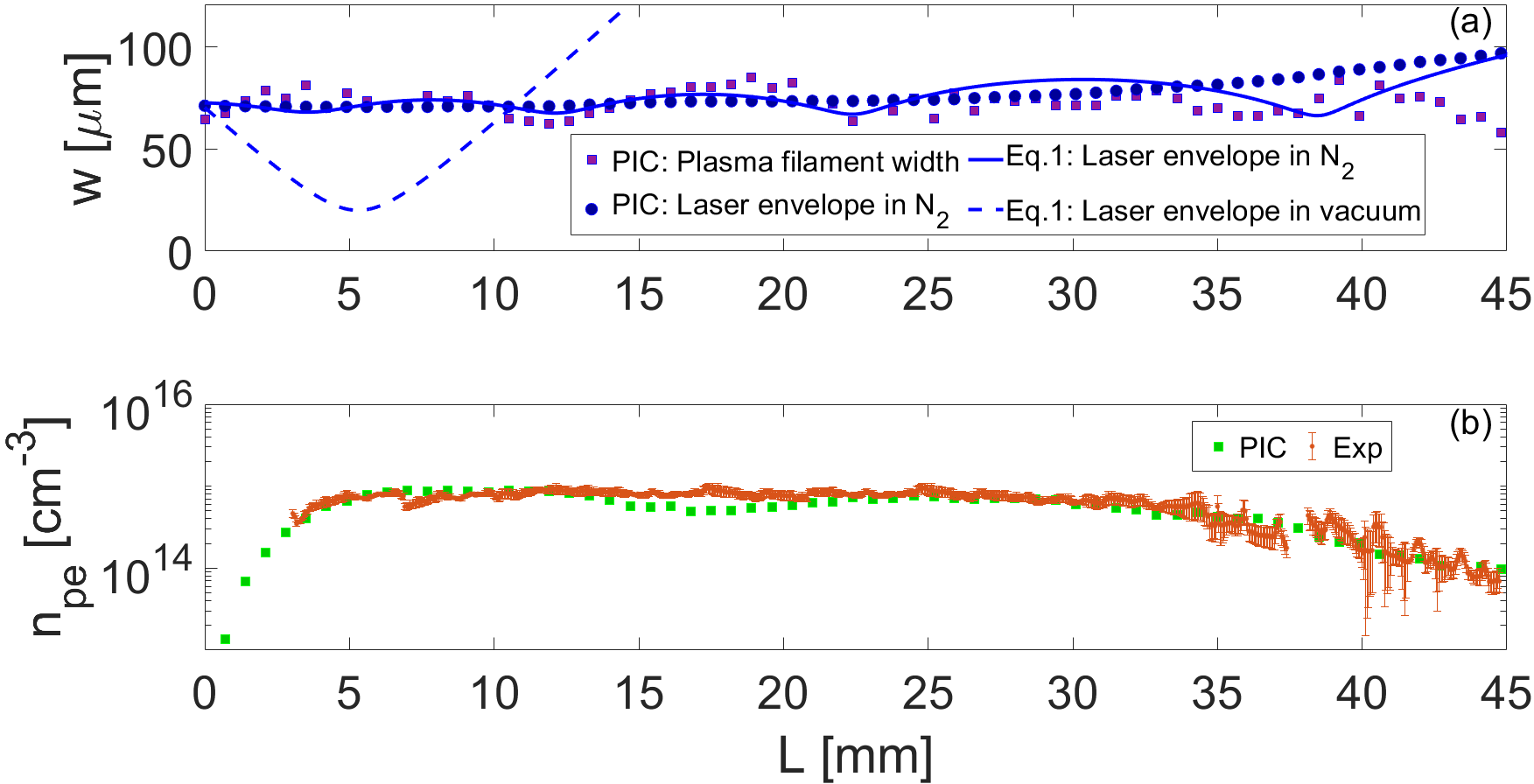}
\caption{\textbf{Laser filamentation simulations.} (a) Laser pulse non-linear propagation retrieved via envelope equation (Eq.~\ref{envelope}) in nitrogen gas (blue continuous line) and in vacuum (blue dashed line); laser pulse non-linear propagation (blue circles) and plasma filament width (violet squares) retrieved from PIC simulation. 
(b) Longitudinal plasma density distribution retrieved from PIC simulations (green squares) and from transverse imaging of the filament (orange circles) as in the inset of Fig.~\ref{fig2}.}
\label{fig1}
\end{figure}

\par The solid blue line in Fig.~\ref{fig1} shows the transverse envelope of a laser pulse with central wavelength $\lambda=800\,$nm, root mean square (rms) transverse size at waist $w_0 = 20\,$\textmu m in vacuum (Rayleigh length $Z_R=\pi w_0^2/\lambda=1.5\,$mm) and Full Width Half Maximum (FWHM) duration $\tau=350\,$fs, interacting with neutral nitrogen gas with density $n_{N}=10^{16}\,$cm$^{-3}$, obtained solving Eq.~\ref{envelope}.
We consider a longitudinal density profile with a plateau from 5 to 35\,mm and smooth ramps at the entrance and exit (similar to the experimental setup, see Methods).
Calculations show that, because of the occurrence of filamentation, the pulse is self-guided for $\sim45\,$mm, corresponding to $\sim 30\,Z_R$, with an average transverse size $w_L\sim70\,$\textmu m.
The dashed blue line shows the corresponding envelope of the pulse when propagating in vacuum.
When filamentation occurs, the focal plane is shifted upstream, and the waist is larger than when propagating in vacuum.

\par We compare the theoretical results with Particle-In-Cell (PIC) numerical simulation results obtained with FBPIC~\cite{LEHE:2016}.
The transverse size of the laser pulse, calculated from the laser electromagnetic fields (blue circles in Fig.~\ref{fig1}), is in good agreement with the theoretical one.
We also extract the width of the plasma channel $w_{FIL}$ (purple squares in Fig.~\ref{fig1}), calculating the rms of the transverse distribution of the plasma electron density of the filament after the pulse has passed. 
The width of the filament is $\sim70\,$\textmu m on average, and it oscillates between 60\,\textmu m and 85\,\textmu m, with a behaviour similar to that of the laser pulse envelope.
The on-axis plasma electron density obtained from the PIC simulations ($n_{pe}~PIC$, green squares in Fig.~\ref{fig1}) reaches an average value of $\sim8\times 10^{14}\,$cm$^{-3}$, along the 3-cm long plateau corresponding to that of the neutral gas density distribution.

\paragraph{Experimental setup} \label{sec2}

\par
\begin{figure}[h]
\centering
\includegraphics[width=\textwidth]{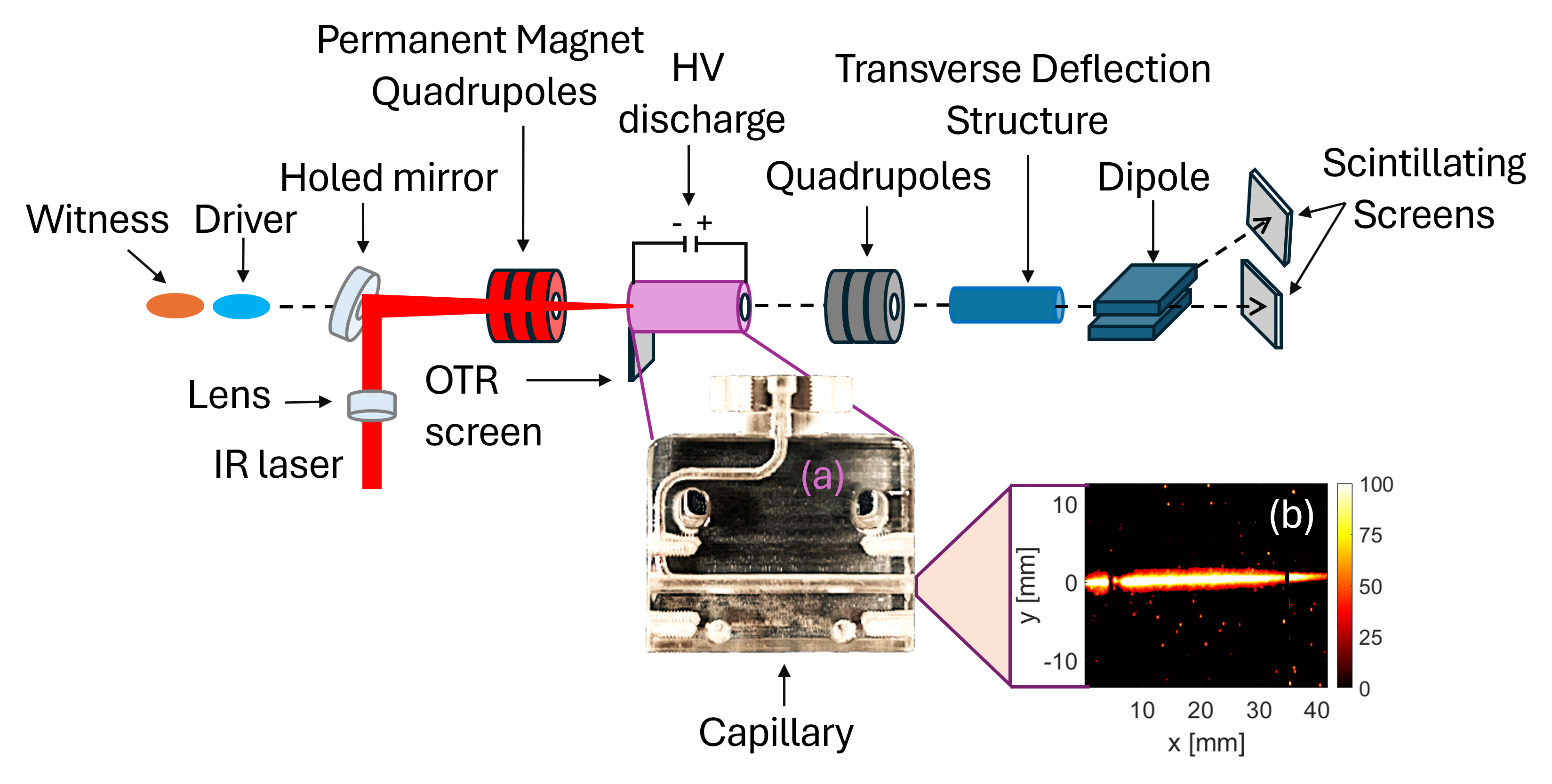}
\caption{\textbf{Experimental setup.} Schematic layout of the SPARC\_LAB linac interaction chamber arranged for the plasma filament-based acceleration experimental campaign. (a) 45$^{\circ}$, single-inlet capillary design. 
(b) Typical filament snapshots taken via the side imaging technique.}
\label{fig2}
\end{figure}

 The experiment took place at the SPARC\_LAB facility at INFN-Frascati National Laboratory~\cite{FERRARIO:2013} (see Fig.~\ref{fig2}). 
The SPARC\_LAB bunker hosts a Ti:Sapphire chirped pulse amplification~\cite{pap16} laser system, which delivers 80~mJ laser pulses at 800~nm central wavelength, with a transform-limited duration of 30~fs at FWHM, and a repetition rate of 10~Hz.
The output of the laser system~\cite{GALLETTI:2022a,pap17} is split into two beamlines.

\par In the first one (linac beamline), the pulse is frequency-doubled to generate electron ($e^-$) bunches illuminating a copper cathode within a radiofrequency (RF) gun. 
Multiple bunches in the same RF bucket can be generated by splitting and delaying the main pulse in replicas with adjustable delay and relative intensity~\cite{pap17,VERRA:2025}.
The bunches are then accelerated ($E\sim97\,$MeV) and longitudinally compressed (sub-picosecond) via velocity-bunching~\cite{SERAFINI:2001,FERRARIO:2010} in the RF linac, and transversally focused with a triplet of permanent magnet quadrupoles (PMQ) at the entrance of a 2-mm-diameter, 3-cm-long, single-inlet dielectric capillary.

\par The second beamline (filament beamline) delivers pulses with $\sim10\,$mJ, 350\,fs (FWHM) and transverse size $\sim15\,$mm at $1/e^2$ intensity at a $2''$ 1-m-focal-length lens, focusing the pulse at the entrance of the capillary, after being reflected by a $45^\circ$ holed mirror through which the electron bunches propagate. 
The laser pulse reaches the capillary entrance approximately 100~ps before the arrival of the electron bunches.

\par The capillary, shown in Fig.~\ref{fig2}~(a), is filled with nitrogen gas through a solenoid valve $1.5\,$ms before the arrival of the laser pulse, and plasma is generated by filamentation.
The inset of Fig.~\ref{fig2}~(b) shows the plasma filament obtained by imaging the recombination light of the plasma onto a multi-channel plate coupled to a CCD camera (side imaging technique).
From these snapshots, we calculate the length of about 45\,mm (Full Width Tenth Maximum, FWTM) and rms transverse size around 70\,\textmu m.
The black regions along the filament are due to the metallic electrodes installed on the capillary for the ionisation with the high-current discharge. 

\par We obtain the longitudinal profile of the plasma by projecting the pixel counts along the vertical axis. 
The trend of the resulting intensity profile (orange dots in Fig.~\ref{fig1}, normalised over the peak value of $n_{pe}~PIC$) is in good agreement with the longitudinal plasma density profile obtained by numerical simulations.
In a previous publication~\cite{GALLETTI:2025} with a dedicated setup, we also directly measured the plasma density and the characteristic decay time with the Stark-broadening technique~\cite{QIAN:2010}.
Since the intensity of the light is correlated with the plasma density, we estimate a density jitter of $6.8\%$ in the plateau region (5-35\,mm, shown in Fig.~\ref{fig1}).


\par We measure the position and transverse distribution of the electron bunches and of the attenuated laser pulse at the capillary entrance using an aluminated silicon screen.
The screen is installed below the entrance of the capillary, and it is inserted in the beam path by moving the capillary vertically with a step motor. 
The optical transition radiation (OTR) emitted by the electrons or the reflected laser pulse is imaged onto the chip of a digital camera. 

\par We characterise the electron bunches' energy and longitudinal distribution using a magnetic spectrometer composed of a triplet of electromagnetic quadrupoles, a transverse deflection structure~\cite{ALESINI:2006} (introducing a head-to-tail vertical correlation to the bunch), a magnetic dipole (dispersing the beam in the horizontal plane) and a scintillating screen.

\par The rms duration of the driver and witness bunches (obtained by calculating the rms of the longitudinal bunch distribution on the screen with TDS turned on) is $\sigma_t=(0.6,0.2) \pm 0.1$~ps, respectively. 
The temporal distance between the two bunches is $\Delta t=2.7 \pm 0.1$~ps.
We calculate the bunch density $n_b=(Q/q)/(2\pi)^{3/2}  c \sigma_t \sigma_r^2$, where $Q$ is the charge per bunch and $\sigma_r$ is the rms transverse size at the plasma entrance (measured with the OTR screen).
The bunches have $Q = (500,50)\pm1\,$pC, and $\sigma_r=(27,20)\pm3\,$\textmu m, thus $n_b=(1.5,0.82)\times 10^{15}\,$cm$^{-3}$.
Hence, we expect the driver bunch to generate wakefields in the blowout regime ($n_b\gg n_{pe}=8\times10^{14}\,$cm$^{-3}$) and the witness bunch to be placed in the accelerating and focusing phase of the wakefields. 
The normalised emittance of the bunches is~$\sim2.9\,$\textmu m (measured with a quadrupole scan technique).

\section{Results}\label{sec3}

\begin{figure}[h]
\centering
\includegraphics[width=\textwidth]{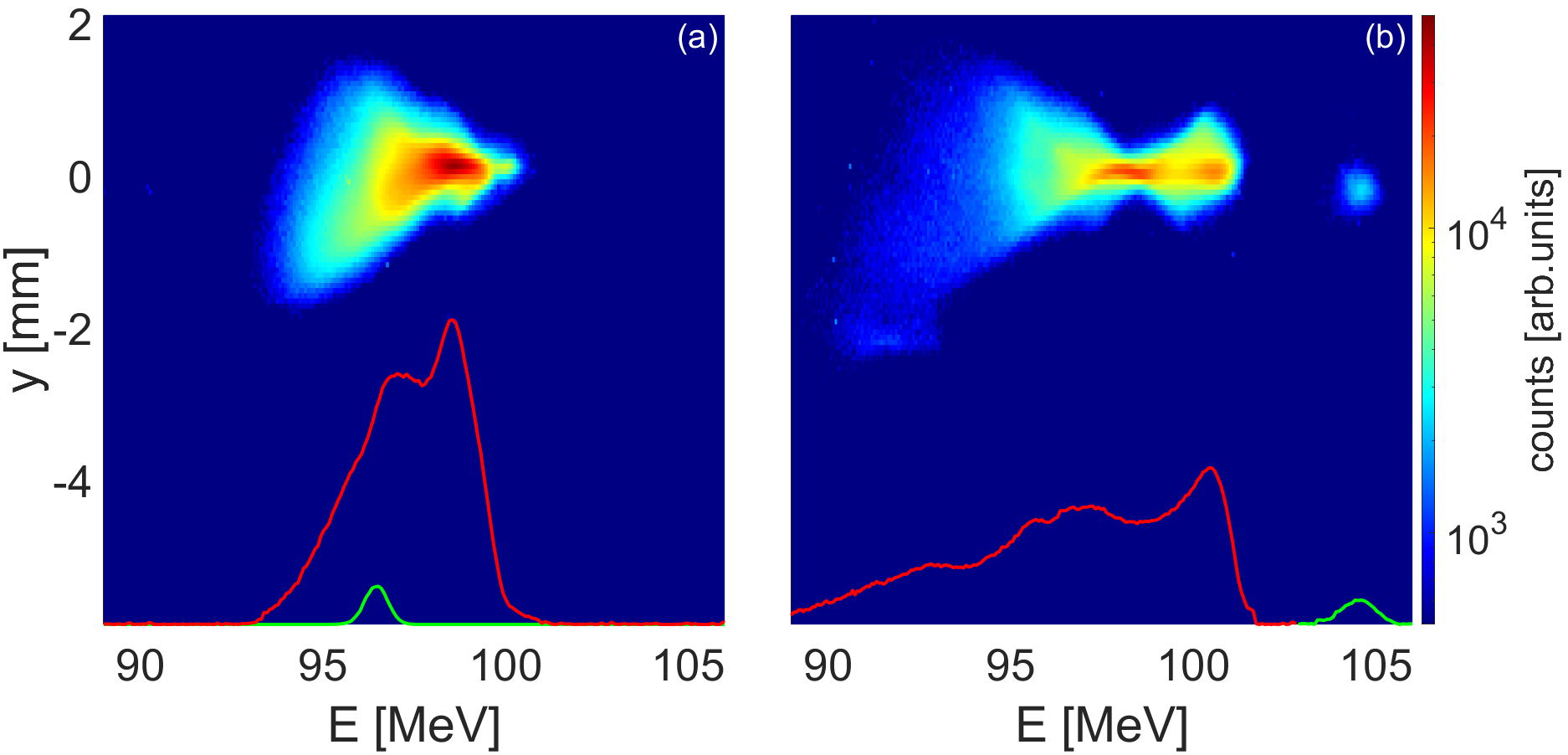}
\caption{\textbf{Experimental energy spectra.} Energy spectrum of driver and witness bunches in the (a) laser-off and (b) laser-on configurations, while gas is present in the capillary. Same colorbar for both plots. 
Solid lines: energy distribution of driver (red) and witness (green).}
\label{fig3}
\end{figure}

\par Figure~\ref{fig3} shows the energy spectra of driver and witness bunches averaged over 100\, consecutive events after propagation through the gas-filled capillary without (a) and with (b) the ionising laser pulse. 
The red and green projections on the energy axis show the energy distribution of the driver and witness bunch, respectively. 
By comparing the spectra, we retrieve a maximum deceleration of the driver $\Delta E^-_{D}\sim4\,$MeV~\footnote{We define the minimum and maximum energy of each bunch as the values where the amplitude of the energy distribution reaches $10\%$ of its own maximum on the low and high-energy sides, while the mean energy is the centroid of the spectral distribution. \label{note1}} and an average acceleration of the witness $\Delta E^+_{W}\sim8\,$MeV, reaching a final energy $E_W=104.5\pm 0.4\,$MeV, when the laser pulse is present. 
This is due to the fact that the laser pulse interacts with the gas, generating the plasma filament within which the electron bunch drives the wakefields. 
Considering a 3-cm long plasma, the average accelerating gradient is $E^+_z\sim 266\,$MeV/m. 
We note that the charge and the relative energy spread of the witness bunch are preserved upon acceleration.

\begin{figure}[h]
\centering
\includegraphics[width=\textwidth]{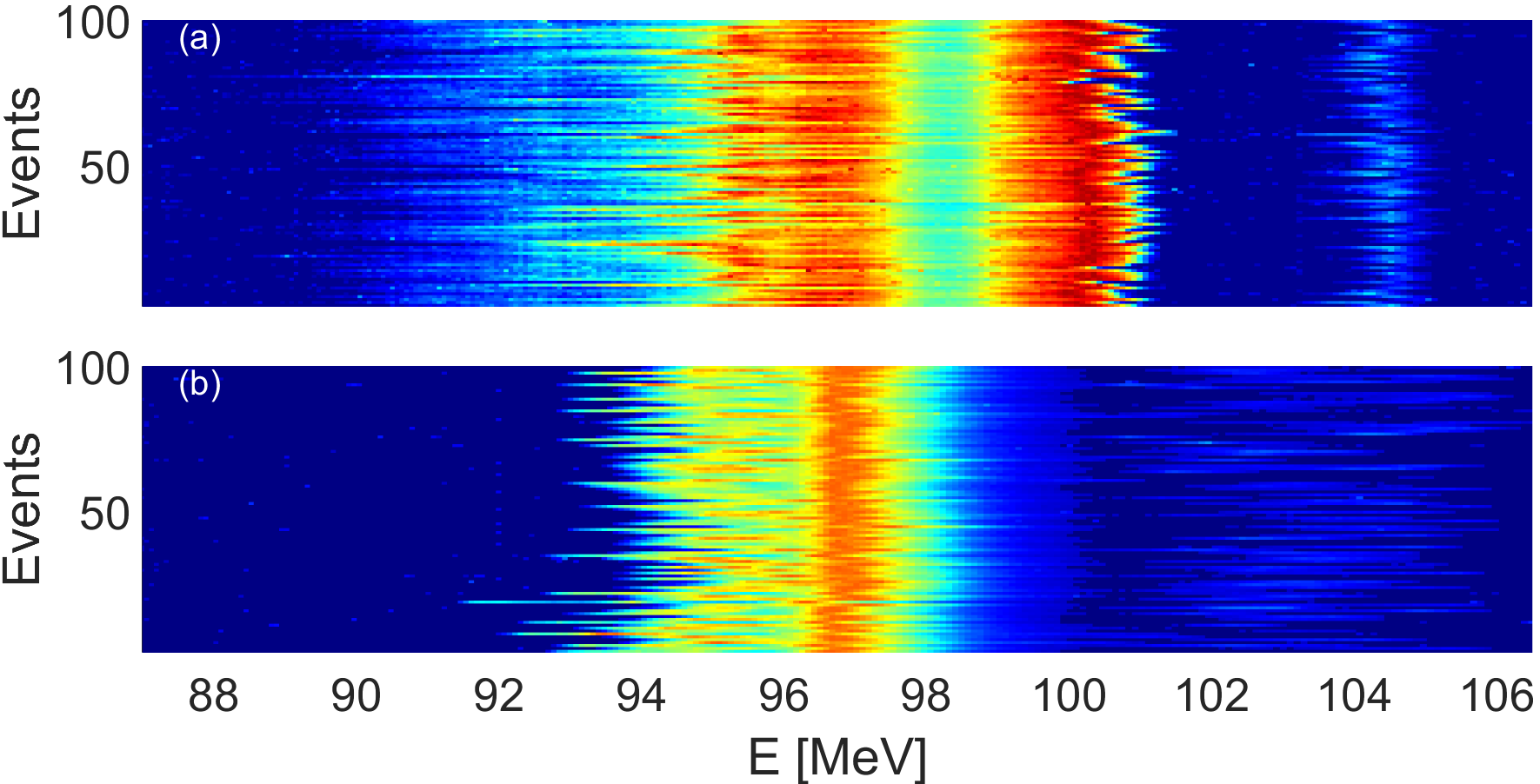}
\caption{\textbf{Waterfall plot of experimental energy distributions.} Energy distribution of driver and witness bunches of 100 consecutive events in the (a) laser-on and in the (b) discharge-on configuration, while gas is present in the capillary with similar initial beam parameters and the same average plasma electron density. The intensity of each plot is normalised with respect to the maximum of (a). }\label{fig4}
\end{figure} 

\par The robustness and reproducibility of the plasma filament generation and of the acceleration process are clearly evidenced by Fig.~\ref{fig4}~(a), where we show the energy distribution of 100\, consecutive events. 
The witness bunch is accelerated in $95\%$ of the events, and the rms energy jitter is $<0.5\%$.

\par An additional demonstration of reliable and repeatable operation of the proposed setup is directly given by a direct comparison with the discharge-based plasma source, which is an established setup for PWFA~\cite{POMPILI:2022,GALLETTI:2022a,GALLETTI:2024}.
Figure~\ref{fig4}~(b) shows that the witness bunch is accelerated in a discharge-based plasma (without additional stabilisation methods~\cite{biagioni2016electron,GALLETTI:2022b}) in only $75\%$ of the events, with rms energy jitter of $\sim1.3\%$, meaning a threefold increase with respect to the filament configuration, with a comparable energy gain.
This is due to the fact that in the discharge configuration, the ionisation process is initiated in a stochastic manner, affecting the plasma density and therefore the acceleration process. 

\par
\begin{figure}[!h]
\centering
\includegraphics[width=\textwidth]{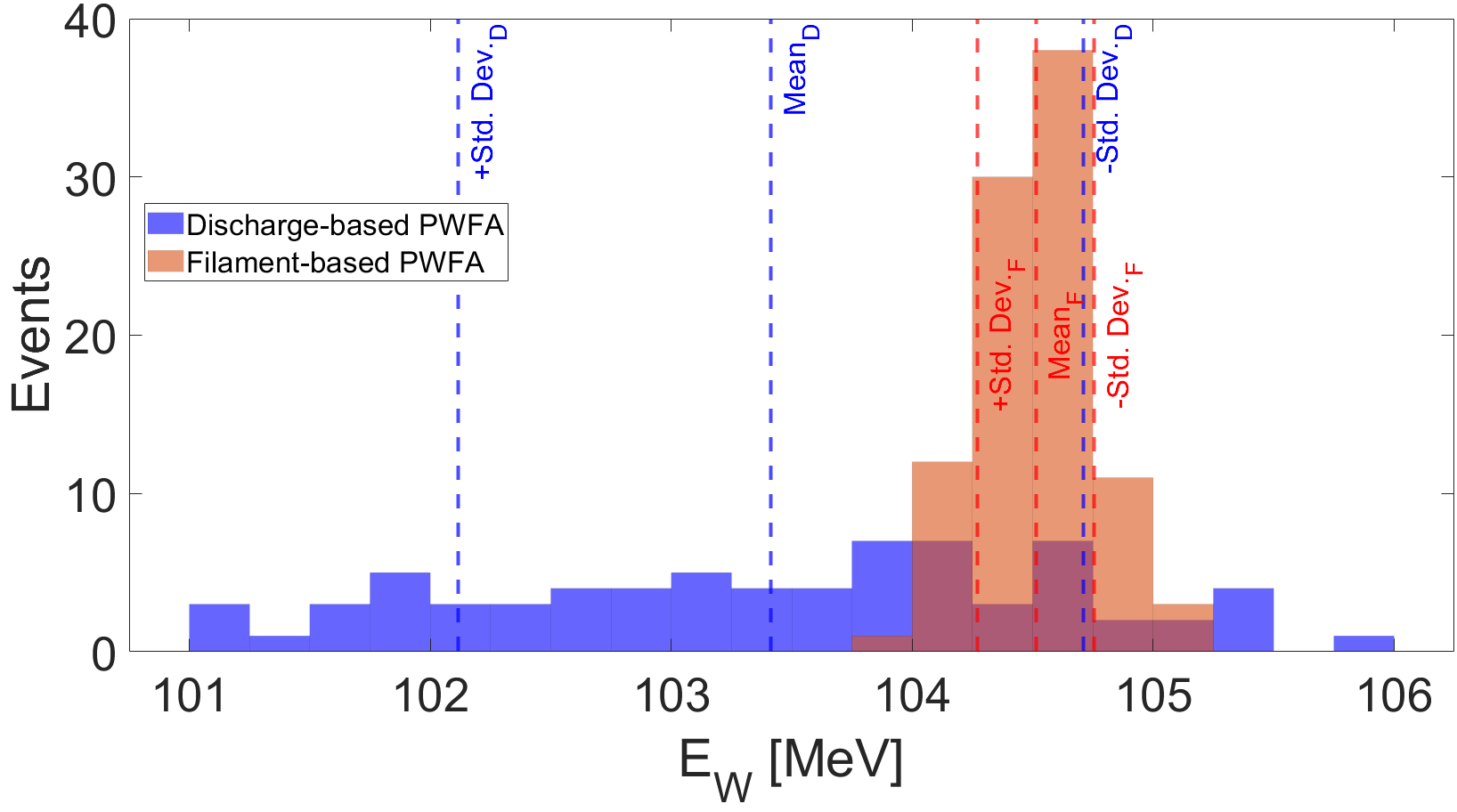}
\caption{\textbf{Comparison between the filament-based PWFA and the discharge-based PWFA.} Histogram of the accelerated witness energy $E_W$ in both filament-based (orange area) and discharge-based (blue area) configurations.}
\label{fig7}
\end{figure}

 We calculate the average energy of the witness bunch $E_W$ of each successful acceleration event. 
The corresponding histograms in Fig.~\ref{fig7} clearly show that the energy variation in the filament case (orange area) is much smaller than in the discharge case (blue area). This is due to the higher reproducibility and inherent synchronisation of the system with the RF linac.


\section{Discussion}\label{sec4}

\begin{figure}[h]
\centering
\includegraphics[width=\textwidth]{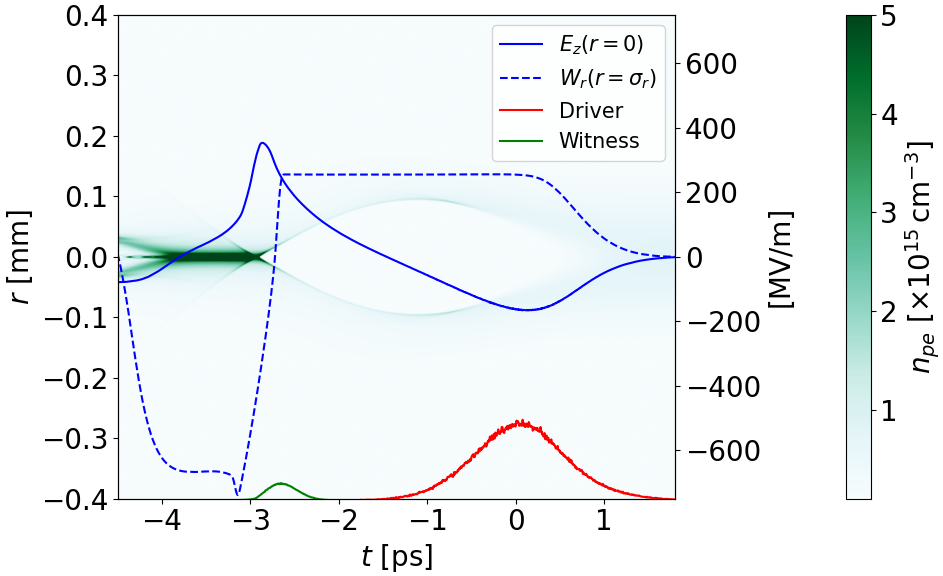}
\caption{\textbf{PWFA PIC simulations.} 
Colourmap: 2D plasma electron density distribution.
Right-hand side axis: blue line, the on-axis ($r=0$) longitudinal ($E_z$) wakefield; blue dashed line, the transverse ($W_r$) wakefields at $r=\sigma_r$. 
The charge density distribution of the driver bunch (red line) and of the witness bunch (green line) is shown in normalised units.}
\label{fig5}
\end{figure}

In the present work, we conducted a proof-of-principle experiment to demonstrate the use of laser filamentation for the generation of plasma stages suitable for PWFA. 
To obtain further insights into the plasma properties, we compare the experimental results with numerical PIC simulations.
The plasma is initialised with a transverse Gaussian distribution with rms $\sigma_p=70\,$\textmu m and on-axis $n_{pe}=8\times10^{14}\,$cm$^{-3}$.
Figure~\ref{fig5} shows the longitudinal (solid blue line) and transverse (dashed blue line) wakefields driven by an $e^-$ bunch with the same parameters as those experimentally measured (driver and witness charge density distributions shown by the red and green lines, respectively, centred at $t=0$, travelling from left to right). 
The plasma density distribution and the properties of the wakefields behind the driver bunch ($t<0$) are those typical of the blowout regime, as expected by $n_b\gg n_{pe}$. 

\par The amplitude of the decelerating wakefields within the driver bunch reaches $\sim150\,$MV/m, while that of the accelerating wakefields within the witness bunch ($t=-2.7\,$ps) reaches $\sim 260\,$MV/m.
Considering a 3-cm long plasma, the amplitude of the longitudinal wakefields is in agreement with the energy loss and gain experimentally measured. 
Hence, we conclude that the interaction of the laser pulse with the gas generates plasma with a longitudinal profile consistent with the one shown in Fig.~\ref{fig1} and transverse Gaussian profile with $\sigma_p = 70\,$\textmu m.

\par We have experimentally demonstrated the generation of an operational PWFA stage satisfying several essential requirements, such as the control of the parameters of plasma, and the reliable and reproducible performance. 
The filamentation process requires significantly less power than with other plasma generation methods, for a given plasma length and density. 
In the configuration described here, a commercial GW-class laser was employed, whereas alternative approaches~\cite{DEMETER:2021,OCONNEL:2006,BIAGIONI:2021,LEEMANS:2006} would require either a TW-class laser, GeV-class relativistic electron beams, or an external high-voltage pulser circuit. 
In fact, in the laser-filament method, the power requirement is due to the partial multi-photon ionisation of the gas; while, in all other methods, more power is required to trigger the thermal heating of the majority of atoms, which in turn produces a sufficient plasma electron density.

\par The filamentation-based configuration results in negligible deposition even at high repetition rates, owing to the relatively low laser energy and the transverse size of the plasma column. 
In the experiment presented here, we measured an energy loss of $\sim2.5$\,mJ by the laser pulse. For a 3-cm-long and 2-mm-wide capillary, this value (assuming conservatively that all the energy is deposited on the wall) corresponds to a thermal load of less than 1 mJ/cm$^2$, which is roughly two orders of magnitude below the damage threshold. Furthermore, in this setup, the capillary serves solely as a gas containment cell, so its diameter could be increased. The only constraint is the beamline pumping system; with an upgrade, the wall load could be reduced even further.
The achievable repetition rate is therefore determined by the laser system. 
Current state-of-the-art lasers with modest pulse energies at the mJ level can operate at repetition rates in the tens of kHz range. 

\par The repetition rate in discharge-based configurations is instead strongly limited by the performance of the high-voltage pulser circuit. 
Stable operation can be sustained routinely at $\sim100\,$Hz for several hours, while operation at kHz would require a complex cooling system.
Additionally, discharge-based configurations exhibit substantially higher energy deposition. 
In a separate experiment, we measured an energy deposition of approximately 100~mJ per discharge in a 3-cm-long capillary, increasing to about 2~J per discharge in a 60-cm-long capillary (as envisioned for the EuPRAXIA$@$SPARC\_LAB project~\cite{assmann2020eupraxia,TDR}). 
Such levels of deposited energy lead to damage to the internal structure of the plasma stage 
unless materials capable of withstanding high thermal load and exhibiting high mechanical strength, but demanding to handle, are employed, thereby enabling operating safely at high repetition rate as in Ref.~\cite{crincoli2025advanced}.
Once the capillary channel is degraded through operation, the resulting plasma profile is altered, leading to a non-optimised acceleration process.

Another advantage of the proposed scheme is that the plasma generated through filamentation is self-synchronised with the electron beam, since the same laser system is used to generate the beam from the photo-cathode. On the contrary, high-voltage plasma-discharges rely on an external pulser unit, which must be precisely synchronised with an appropriate timing system~\cite{GALLETTI:2022b}.
As a result, laser-based plasma formation offers improved reproducibility and control over plasma parameters, resulting in a threefold reduction of the energy jitter and in an improved success rate (see Fig.~\ref{fig4} and Fig.~\ref{fig7}).

\par Moreover, the plasma filament dimensions and density required for an optimal PWFA stage are determined solely by the laser and gas parameters, namely the proper matching of laser energy, pulse duration, waist, and gas type and density, as derived from Eq.~\ref{envelope}. 
We note that a longer plasma could be generated with the same laser pulse as in this experiment, using a longer capillary, whose length is limited in our facility by the geometric dimensions of the interaction chamber. 

\par Finally, this technique enables the generation of plasma channels with adjustable transverse dimensions.
For example, it is possible to obtain plasma channels much narrower than the driver bunch, enabling ion-channel applications~\cite{ionchannel}, since all plasma electrons would be expelled from the plasma and are not able to restore their initial condition,
as envisioned for future plasma-based undulators.
Finite-width plasma stages may also be suitable for positron acceleration, thanks to the extended accelerating and (transversely linear) focusing region for positively charged particles~\cite{Diederichs:2019} (see Fig.~\ref{fig5}, $-4.4\,$ps $<t<-4.0\,$ps), which are required for next-generation plasma-based lepton collider~\cite{GESSNER:2025}.

\par In conclusion, we conducted a proof-of-principle experiment, supported by numerical simulations and theoretical results, demonstrating that laser filamentation of ultrashort pulses can generate plasma channels suitable for PWFA.
This approach offers significant advantages, enabling future operation of high-gradient plasma stages at high repetition rate while maintaining minimal energy deposition on the gas-cell walls.
This is therefore aligned with the requirements defined in the EuPRAXIA scientific case~\cite{TDR}, which foresees the use of a high-repetition-rate, 60-cm-long plasma with a density in the $10^{15}$~cm$^{-3}$ range, capable of sustaining GV/m accelerating fields while preserving the high brilliance of externally injected beams. 

    
\section{Methods}\label{sec5}

\subsection{Laser system}

\par The SPARC\_LAB photocatode laser system is based on a commercial Ti:Sapphire chirped pulse amplified laser (ARCO~\cite{Amplitude}) delivering $80\,$mJ, 30\,fs transform-limited duration, $800\,$nm central wavelength (IR) pulses at 10 Hz repetition rate.

\par The main high-energy 10 Hz laser beamline generates pulses with linear polarization, beam quality factor $M^{2}\sim 1.2$, picosecond contrast of $10^{-5}$ for $<-5$ ps  - $10^{-6}$ for $< -10$ ps - $5\times10^{-7}$ for $< -50\,$ps, and pointing stability of about $< 10\,$\textmu rad.

\par The main beamline is split into two auxiliary beamlines: one beamline is used to generate the electron bunches, namely the driver and the witness bunches; while the second one is used to generate the plasma filament for the PWFA stage. 

\subsection{Generation of electron bunch train}

In the linac beamline, the pulse frequency is doubled (VIS) with $\beta$-BBO non-linear crystals and shaped in energy, transverse and longitudinal dimensions. The harmonic conversion delivers up to $4\,$mJ, with a nominal rms stability of about $1.5\%$.

The split and delay system is composed by custom made polarizing beam splitters (PBS) that reflect vertical polarisation and transmit horizontal polarisation. 
Motorised zero-order half-wave plates rotate the polarisation of the incoming pulse to control the intensity of each VIS replica.  Translation stages in each line allow for control of the delay between the pulses. Two motorised irises, positioned at the same distance from the first PBS, control the transverse dimensions of the two pulses. 

\par The VIS pulse train is transported to the copper photocathode, where imaging of the irises is performed to obtain the top-hat transverse profile needed to optimise the photoelectron generation.  
About $40\%$ of the pulse is reflected by the cathode, and it is imaged on a camera used for transverse alignment and measurement of the pulse size, and as a monitor for copper surface deterioration. 

\begin{figure}[h]
\centering
\includegraphics[width=0.9\textwidth]{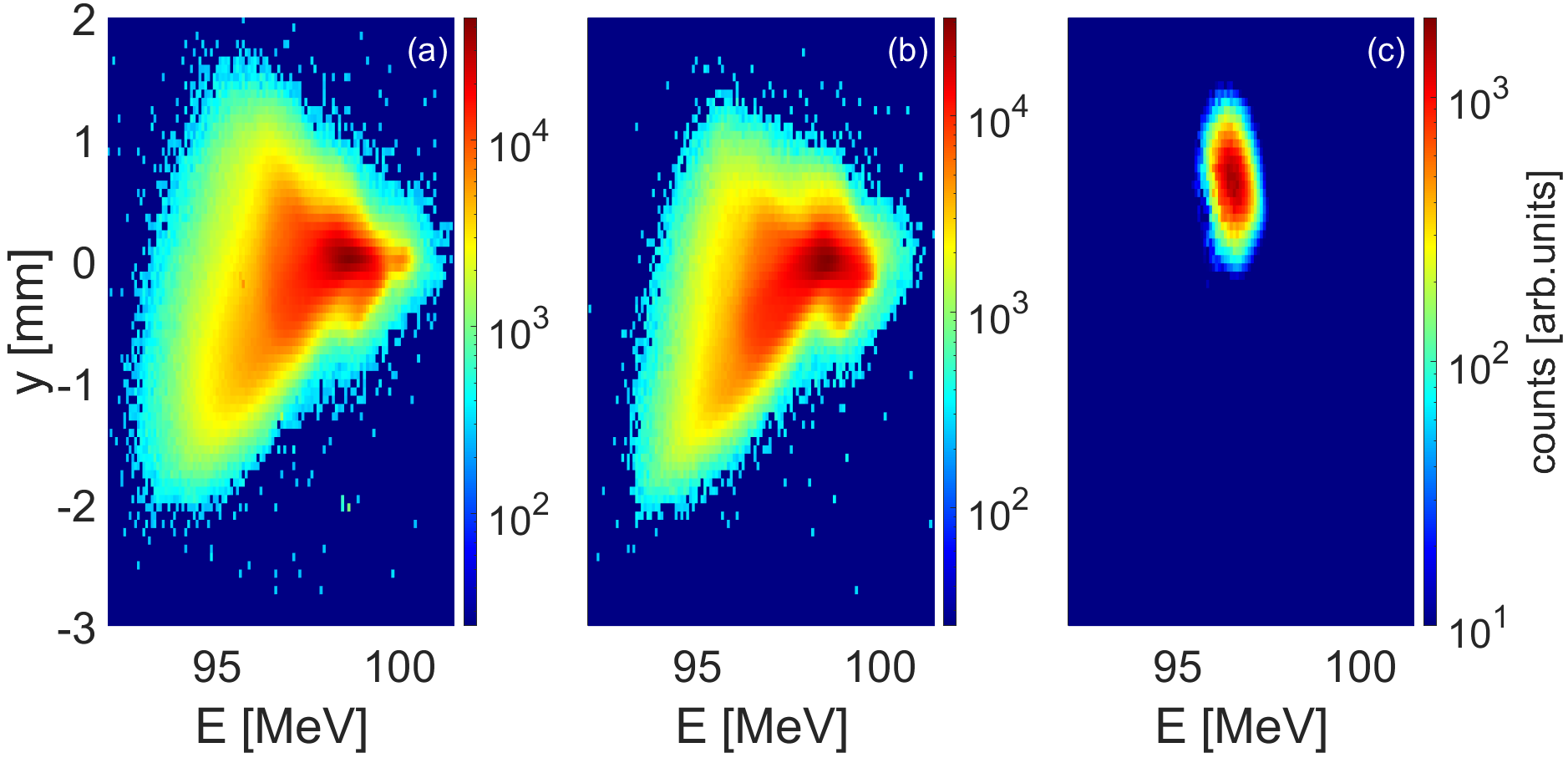}
\caption{\textbf{Experimental energy spectra.} Energy spectrum of driver and witness bunches (a), and of only driver (b). The witness bunch energy spectrum (c) is retrieved by subtracting (b) from (a). The witness has an energy of 96.5 MeV with a relative energy spread $<1\%$.}\label{fig6}
\end{figure}

\par The electron bunches are accelerated to $E\sim97\,$MeV, as shown in Fig.~\ref{fig6}, adopting three RF travelling-wave structures: two S-band (2.85\,GHz), where the first one is working in the velocity-bunching configuration~\cite{SERAFINI:2001, FERRARIO:2010} for bunch compression, and one C-band (5.71\,GHz).
The total charge is measured with an integrating current transformer at the exit of the RF gun and downstream of the plasma. The emittance of each bunch is measured with the quadrupole scan technique at the end of the linac. 

\subsection{Laser-filament plasma source}

\par
\begin{figure}[ht]
\centering
\includegraphics[width=\textwidth]{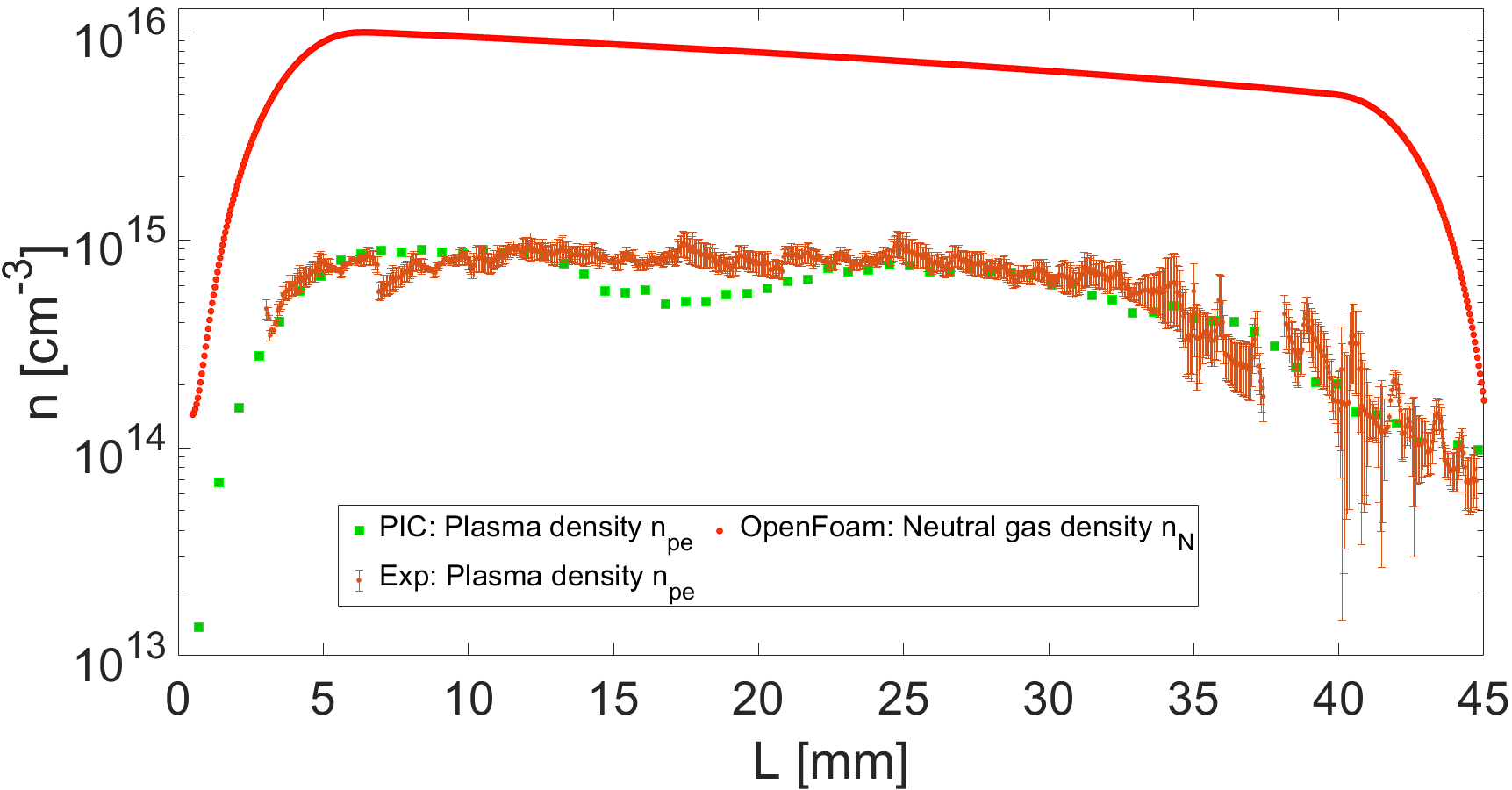}
\caption{\textbf{Longitudinal density distributions.} Longitudinal neutral density distribution retrieved from OpenFoam simulations (red circles), longitudinal plasma density distribution retrieved from PIC simulations (green squares) and from transverse imaging of the filament (orange circles) as in the inset of Fig.~\ref{fig2}.}
\label{fig8}
\end{figure}

 The capillary shown in Fig.~\ref{fig2}~(a) was specifically designed for this experimental campaign to optimise gas injection and longitudinal neutral density distribution (shown in Fig.~\ref{fig8}) and, in turn, ensure optimised matching with the laser pulse.

The capillary (and the PMQ, electron focusing system) is installed in a vacuum chamber connected to the linac with a windowless, three-stage differential pumping system. 
The IR laser pulse is transported through the second beamline to the vacuum-chamber window, which is positioned at 90$^{\circ}$ with respect to the longitudinal axis. 
Upstream of the window, a 2'' 1-m focal-length lens mounted on a motorised translation stage is used to accurately position the focal plane at the capillary entrance. 
The laser pulse is then re-injected on axis, corresponding to the capillary longitudinal axis, using a 45$^{\circ}$ holed mirror, which reflects approximately 90\% of the laser energy while allowing 100\% of the electron-bunch distribution to pass through the central aperture.

\par The experiment is conducted using a 3-cm-long, 2-mm-diameter single-inlet dielectric capillary. Nitrogen gas is injected through the inlet connected to a solenoid valve located 5\,cm from the capillary that remains open for 3\,ms. 

\par We measure the time of arrival of the electron bunches with respect to the peak of the IR laser pulse with the analog signal generated by a beam position monitor upstream of the vacuum chamber with a GHz scope. The timing jitter between the bunch and the IR laser is negligible, being generated by the same main laser system. As a result, the plasma density at the electron bunches' arrival time is constant from event to event.
By varying the delay between the laser pulse and the electron bunches' arrival time, we vary $n_{pe}$, due to recombination of the plasma.

\par The plasma density is retrieved with the side imaging technique calibrated with the Stark-broadening-based diagnostics. The latter consists of measuring the spectral width of the $H_\alpha$ and $H_\beta$ Balmer lines using a mixture of 95\% nitrogen and 5\% hydrogen gas, with an imaging spectrometer coupled to an intensified digital camera, as in Ref.~\cite{GALLETTI:2025}.
The plasma electron density is essentially constant along the capillary, but non-uniformities are present at the two extremes as the gas flow expands in the vacuum chamber, as shown in Fig.~\ref{fig2}.

\subsection{Side imaging diagnostics}

\par Using a CCD coupled with a Hamamatsu C9547-01 intensifier (ICCD), we image the light emitted by the plasma channel from the side~\cite{pap7:21}. 
The ICCD camera is positioned perpendicularly to the pulse propagation axis, and the plasma emitted light signal is collected and imaged onto the ICCD detector by using a Sigma 180~mm MACRO f$\/$~2.8 \textit{EX-DG-OS-HSM} objective. 

\par A band-pass filter coupled with a 0$^\circ$ incident 800~nm dielectric mirror is placed in front of the camera to detect the light emission from the filament while rejecting the scattered light from the IR laser and all other emissions. 

\par The section of the plasma column covered by the field of view of the ICCD is about 5$\times$3~cm$^2$. 
The ICCD gate duration is set to 5\,ns with an available scanning delay with respect to the laser pulse arrival time.

\par Adopting the side-imaging technique, we directly measure the transverse size and length of the plasma column, as shown in Fig.~\ref{fig1}. 
Moreover, the plasma-emitted light is directly related, through the Einstein coefficient, to the number of ions in the excited state, which, up to a calibration factor, provides a measure of the plasma density~\cite{pap7:21,pap7:35}. 
To determine the plasma density quantitatively, the emitted light intensity is calibrated using the Stark broadening technique.

\subsection{PWFA in a discharge-generated plasma stage}
To benchmark the filament-based plasma stage, we performed a dedicated experimental campaign using a discharge-based plasma source (adopting the same capillary), operating in a driver–witness configuration with beam parameters matched to those adopted for the filament case. This ensured a consistent comparison between the two plasma generation schemes under otherwise equivalent conditions.

\begin{figure}[h]
\centering
\includegraphics[width=\textwidth]{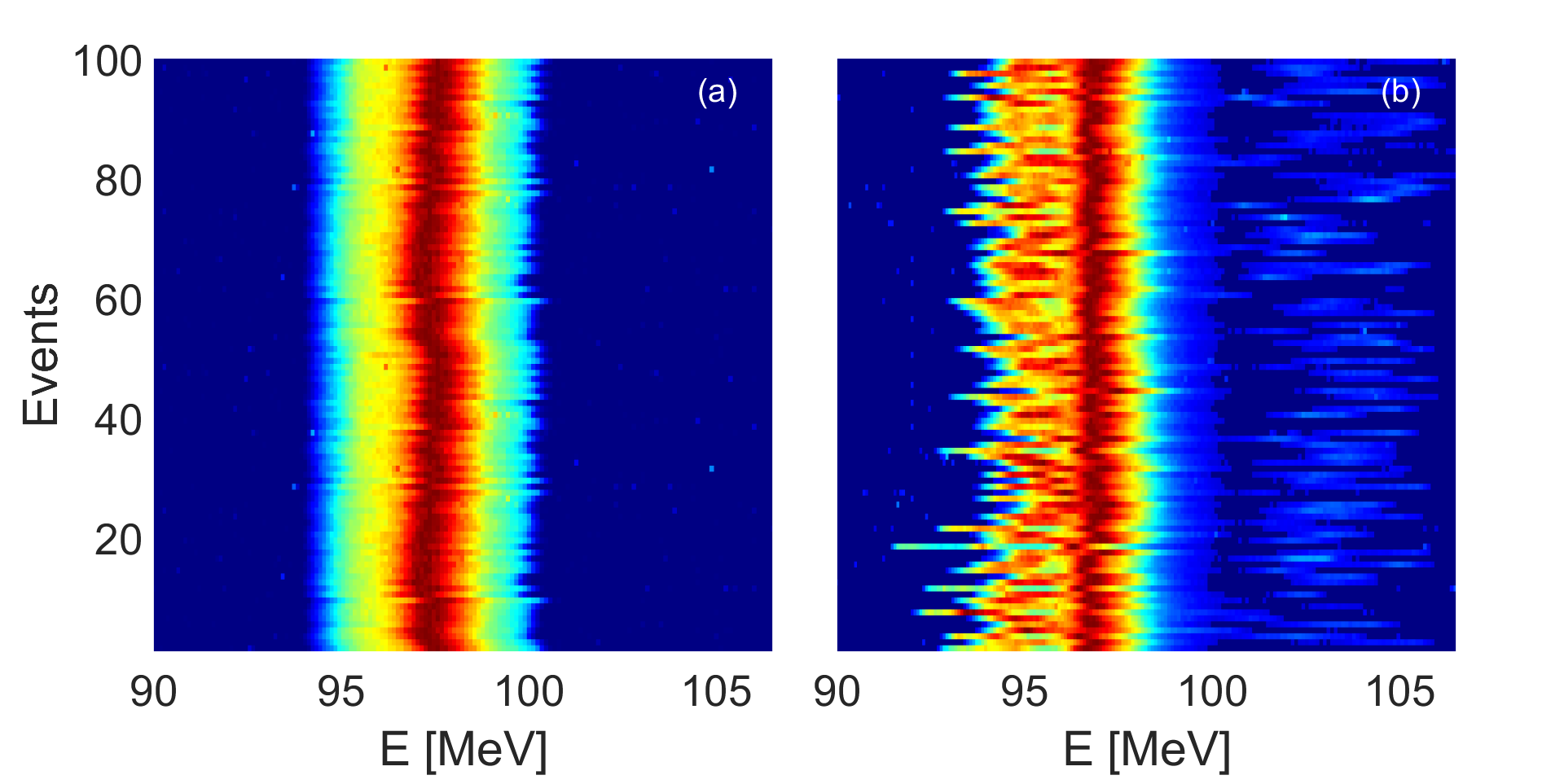}
\caption{\textbf{Waterfall plot of experimental energy distributions.} Energy distribution of driver and witness bunches of 100 consecutive events in the a) discharge-off and b) discharge-on configurations, while gas is present in the capillary.  The intensity of each plot is normalised with respect to its own maximum. }\label{fig9}
\end{figure}

Figure~\ref{fig9} reports the energy distributions of the driver and witness bunches measured over 100 consecutive shots after propagation through the gas-filled capillary, both in the absence (a) and in the presence (b) of a high-current discharge (7~kV, 233~A) with similar beam parameters as in Fig.~\ref{fig4} of the main text. 
For the driver bunch, a maximum energy loss of $\Delta E^-_{D}\sim2.3$~MeV is observed\footref{note1}. The witness bunch, initially at an energy of approximately 97 MeV as shown in Fig.~\ref{fig9}~(a), reaches a final energy of $E_W=103.6\pm1.26$~MeV after propagation through the discharge plasma, as shown in Fig.~\ref{fig9}~(b), corresponding to an average energy gain of $\Delta E^+_{W}\sim7$~MeV. Assuming uniform acceleration over the 3-cm-long plasma channel, the inferred maximum accelerating and decelerating gradients are $E^+_z\sim233$~MeV/m and $E^-_z\sim80$~MeV/m, respectively. In this configuration, despite charge preservation, the accelerated witness bunch exhibits a relative energy spread of approximately $4\%$, nearly an order of magnitude larger than that obtained with the filament-based configuration, indicating a significant degradation of energy spread during acceleration.

\subsection{Particle-In-Cell simulations}

Numerical simulations were performed with the Particle-In-Cell code FBPIC~\cite{LEHE:2016}\footnote{For the code documentation, see \url{https://fbpic.github.io/index.html}}, a quasi-3D PIC code, which uses cylindrical symmetry to save computational time and azimuthal Fourier decomposition to add 3D features to plasma and laser evolution.

\subsubsection{Laser evolution and filament generation simulations}

\par Two azimuthal modes were set in the performed laser-evolution and filament-generation simulations.
The Gaussian laser is initialised with the same parameters as those experimentally measured, meaning $a_0 = 0.01$ with its peak intensity placed at $z=250\,$\textmu m and its evolution is computed using the OpenPMD\footnote{For documentation see~\url{https://openpmd-viewer.readthedocs.io/en/latest/}.} library.
The simulation box is composed of $4500\times300$ grid cells (in a $z-r$ plane) moving with the speed of light velocity, with space resolution $dz = 0.1$\,\textmu m, $ dr = 0.5$\,\textmu m and time step resolution $dt = 0.33\,$fs. 
The laser interacts with a 3-cm long column of neutral nitrogen gas of $n_N = 10^{16}\,$cm$^{-3}$ (adopting the experimentally retrieved distribution, as shown in Fig.~\ref{fig2}), described by $16$ particles-per-cell.
Ionisation is modelled using the Ammosov-Delone-Krainov (ADK) theory~\cite{ammosov1986ADKmodel}.
We calculate the value of the plasma electron density $150\,$\textmu m behind the laser peak intensity ($z=100\,$\textmu m).

\subsubsection{PWFA stage simulations}

\par One azimuthal mode was set in the performed PWFA stage simulations. 
The two particle bunches are initialised in a 6D phase space with the same parameters as those experimentally measured. 
Each bunch is composed of $10^6$ constant-weighted macro-particles. 
The simulation box is composed of $3000\times300$ grid cells (in a $z-r$ plane) moving with the speed of light velocity, with space resolution $dz = 1.83$\,\textmu m, $ dr = 1.3$\,\textmu m and time step resolution $dt = 2.8\,$fs. 
The plasma is initialised in the box at longitudinal position $z=0\,$mm (before that the bunches propagate in vacuum) until the end of the simulations ($z=5.4\,$mm). The plasma transverse distribution is Gaussian with a $\sigma_p=70$ \textmu m and an on-axis peak density of $8\times10^{14}\,$cm$^{-3}$.  

\bmhead{Acknowledgements} We thank I.~Balossino,  M.~Bellaveglia, F.~Cardelli, C.~Di Giulio, L. Piersanti, A.~Vannozzi, A.~Michelotti, F.~Anelli, M.~Ceccarelli, G.~Grilli, M.~Martini, G.~Luminati, and L.A. Rossi for the technical support; M.~Del Giorno for the maintenance and operation of the laser system; V.~Lollo and M.~Zottola for the experimental chamber installation; D.~Pellegrini, G.~Grilli, and T.~De Nardis for the HV discharge pulser design, and all the machine operators involved in the experimental run. Technical support of the Department of Physics of Tor Vergata University of Rome, in particular F.~Bonaccorso for computational resources utilisation support, is gratefully acknowledged.

\bmhead{Funding} This work has received funding from the European Union’s Horizon Europe research and innovation programme under Grant Agreement No. 101079773 (EuPRAXIA Preparatory Phase).

\bmhead{Conflict of interest} The authors declare no conflict of interest.

\bmhead{Data availability} Additional inquiries about the codes should be directed to M.G. (mario.galletti@lnf.infn.it).

\bmhead{Code availability} Additional inquiries about the codes should be directed to M.G. (mario.galletti@lnf.infn.it).

\bmhead{Author contribution} M.G., R.P. and A.Z. conceived the experiment. M.G., R.P., L.V. and G.D.P planned and managed the experiment with inputs from all the co-authors. M.G., L.V., F.S. and F.V. managed the photo-cathode laser system and filament beamline. L.C., R.D. and A.B. designed and characterised the capillary system. M.G., M.C., R.D., A.B., F.S., L.V. and R.P. carried out the experimental campaign. M.G. and M.C. performed the data analysis. M.G., L.V. and G.P. provided numerical simulations. A.Z. and M.F. supervised the project. The manuscript was written by M.G. and L.V. with assistance from R.P. and A.Z., and inputs from all the co-authors.  All authors extensively discussed the results and reviewed the manuscript.

\bibliography{Plasma_filament}


\end{document}